\documentclass[prl,reprint,showpacs]{revtex4-1}
\usepackage{amssymb,amsmath}
\usepackage{subfigure}
\usepackage{graphicx}
\usepackage{mathptmx}
\usepackage{enumerate}
\usepackage{CJK}
\usepackage{color}
\usepackage[breaklinks=true,colorlinks=true,bookmarks=false,urlcolor=blue,
citecolor=blue,linkcolor=blue,hypertexnames=false]{hyperref}
\usepackage{fancyhdr}

\def\unit#1{\mathord{\thinspace\rm #1}}

\definecolor{ur}{rgb}{0.071,0.388,0.678}

\fancypagestyle{plain}{
\fancyhf{}
\fancyfoot[C]{\small \thepage}
\fancyhead[C]{\small Published version: \href{http://journals.aps.org/prl/abstract/10.1103/PhysRevLett.118.066801}{Phys.\ Rev.\ Lett.\ \textbf{118}, {066801} (2017)}}
}

\begin{document}

\begin{CJK*}{UTF8}{}

\title{Creating and Steering Highly Directional Electron Beams in Graphene}

\author{Ming-Hao Liu (\CJKfamily{bsmi}{劉明豪})}
\email{minghao.liu.taiwan@gmail.com}
\affiliation{Institut f\"ur Theoretische Physik, Universit\"at Regensburg, D-93040 Regensburg, Germany}

\author{Cosimo Gorini}
\affiliation{Institut f\"ur Theoretische Physik, Universit\"at Regensburg, D-93040 Regensburg, Germany}

\author{Klaus Richter}
\affiliation{Institut f\"ur Theoretische Physik, Universit\"at Regensburg, D-93040 Regensburg, Germany}

\date{\today}

\begin{abstract}

We put forward a concept to create highly collimated, non-dispersive electron beams in pseudo-relativistic Dirac materials such as graphene or topological insulator surfaces. Combining negative refraction and Klein collimation at a parabolic \textit{pn} junction, the proposed lens generates beams, as narrow as the focal length, that stay focused over scales of several microns and can be steered by a magnetic field without losing collimation. We demonstrate the lens capabilities by applying it to two paradigmatic settings of graphene electron optics: We propose a setup for observing high-resolution angle-dependent Klein tunneling, and, exploiting the intimate quantum-to-classical correspondence of these focused electron waves, we consider high-fidelity transverse magnetic focusing accompanied by simulations for current mapping through scanning gate microscopy. Our proposal opens up new perspectives for next-generation graphene electron optics experiments.

\end{abstract}


\maketitle

\end{CJK*}

\thispagestyle{plain}

The recent development of high-mobility graphene samples, showing ballistic dynamics of Dirac fermions over distances of several microns, has spurred an impressive renewal of interest in coherent charge transport and interference phenomena in graphene. Accordingly, over the past few years, novel transport features of electrons in ballistic single-layer graphene have been reported, such as Fabry-P\'erot interference \cite{Young2009,Rickhaus2013,Grushina2013,Handschin2016}, signatures of the Hofstadter butterfly in exfoliated graphene on hexagonal boron nitride (hBN) \cite{Ponomarenko2013,Dean2013} and in epitaxial graphene grown on hBN \cite{Yang2016}, snake states along \textit{pn} junctions \cite{Taychatanapat2015,Rickhaus2015}, gate-defined electron wave guides \cite{Rickhaus2015a,Kim2016}, negative refraction \cite{Lee2015}, ballistic Josephson junctions \cite{Calado2015,Shalom2016} and transverse magnetic focusing \cite{Taychatanapat2013,Calado2014,Morikawa2015,Bhandari2016,Lee2016}. Such experimental achievements \footnote{Although most of such experiments were done on exfoliated graphene samples, ballistic transport has also been demonstrated using graphene grown by chemical vapor deposition \cite{Calado2014,Banszerus2016}.}, together with improved numerical techniques allowing for a one-to-one modeling of the measurement setups, put closer within reach true ``optics'' or even ``quantum optics'' applications in graphene. Despite such a stunning progress, however, decent control of electron wave propagation in graphene is still limited. In particular, the lack of a source or mechanism for providing narrow and well collimated beams still prevents graphene electron optics from fully taking advantage of its optics-like electronic characteristics. 

\begin{figure}[b]
\includegraphics[width=\columnwidth]{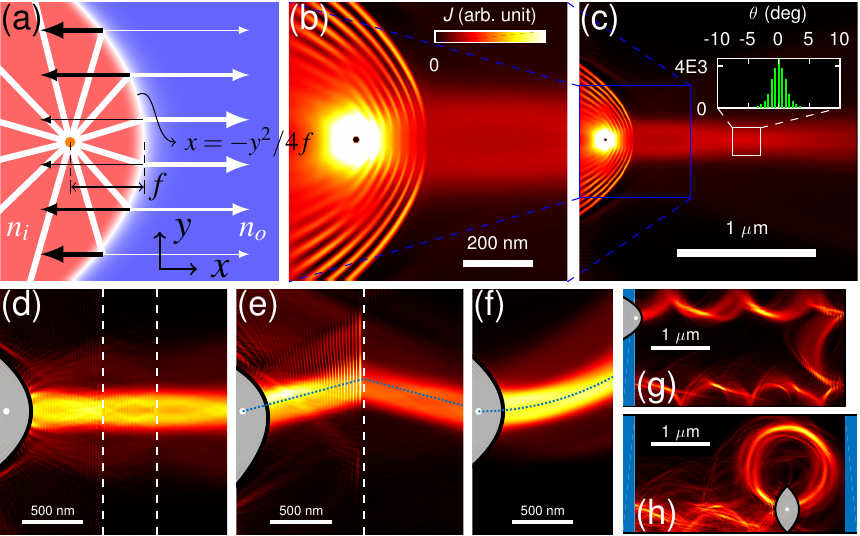}
\caption{(a) Schematic of the lensing apparatus composed of a point-like source at the focal point of a parabolic interface separating two regions with densities $n_i$ and $n_o=-n_i$. (b) An example of the probability current density distribution for design (a),  zoomed-in from (c). Inset in (c): angle (with respect to $x$-axis) distribution of the current density analyzed for the white box area. Gallery of panels (d)--(h) showing the electron beam versatility: (d) nearly-perfect Klein tunneling, (e) negative refraction, (f) bending in a perpendicular magnetic field $B$, (g) ``skipping beam'' in $B$-field along the edge of a graphene cavity, (h) beam (from double lens) bent by the $B$-field to form a full cyclotron orbit. Parameters used: Focal length $f=200\unit{nm}$ in (b--f,h) and $100\unit{nm}$ in (g); carrier density $n_o=6\times 10^{11}\unit{cm^{-2}}$ in (b)--(g) and $7\times 10^{11}\unit{cm^{-2}}$ in (h) (Fermi wavelength $\approx 46\unit{nm}$ and $42\unit{nm}$, respectively); magnetic field $B=40\unit{mT}$ in (f) and $B=150\unit{mT}$ in (g,h). Vertical white dashed lines in (d)/(e) mark an additional potential barrier/step with density $n_i$. \label{fig1}}
\end{figure}

Motivated by the recent realization of point contacts in hBN-encapsulated graphene \cite{Handschin2015}, here we propose and apply a conceptually simple but efficient electron collimator for point sources in graphene, exploiting negative refraction unique to Dirac materials. Contrary to usual Klein collimation \cite{Cheianov2006} or supercollimation in superlattices \cite{Park2008}, we consider a parabolic \textit{pn} junction with a point-like source located at its focal point; see Fig.~\ref{fig1}(a). Paraboloidals have a wide variety of applications, from flashlight reflectors to radiotelescope antennas \cite{Hecht2015}, where either a wave emitted from a point source is turned into a plane wave by specular reflection [black arrows in Fig.~\ref{fig1}(a)], or vice versa. For a point source of waves to \emph{refract} toward an identical direction parallel to the parabola axis [white arrows in Fig.~\ref{fig1}(a)], on the other hand, the refraction indices inside and outside the parabolic \textit{pn} junction must be of opposite sign, provided that the point source is located at the focal point. In graphene, the role of the refraction index is played by the Fermi energy relative to the Dirac point, and hence the carrier density relative to the charge neutrality point. Thus a parabolic electron lens with individually controllable inner and outer carrier densities $n_i$ and $n_o$ can be realized by electrical gating. 

Most notably, when $n_i = -n_o$, the refracted electron waves are expected not only to collimate into a unidirectional wave, but also to concentrate in intensity in a narrow range around the parabola axis due to Klein collimation \cite{Cheianov2006}, i.e., the perfect transmission probability across the \textit{pn} junction at normal incidence, known as the Klein tunneling \cite{Klein1929,Katsnelson2006}, rapidly decreases with increasing angle of incidence. This combined effect generates a highly directional electron beam with width of the order of the parabola focal length $f$. This is illustrated in Fig.~\ref{fig1}(b) by the local probability current density for $f=200\unit{nm}$. Throughout the paper, we refer to the parabolic \textit{pn} junction with densities $n_i=-n_o$ combined with a point-like source at its focal point as the lensing apparatus.

The probability current density images are obtained by the real-space Green's function method in the tight-binding framework \cite{Datta1995}. On site $\mu$ at $(x_\mu,y_\mu)$, the local probability current density at energy $E$ is given by the sum over the bond current vectors to the nearest neighboring sites,
\begin{equation}
\mathbf{J}(E;x_\mu,y_\mu) = \sum_{\nu\in\text{n.n.}} J_{\mu\rightarrow\nu}(E)\mathbf{e}_{\mu\rightarrow\nu}\ ,
\label{eq Jvec}
\end{equation}
with $\mathbf{e}_{\mu\rightarrow\nu}$ the unit vector pointing from $\mu$ to $\nu$, and
\begin{equation}
J_{\mu\rightarrow\nu}(E) = \frac{v_F}{4\pi S} [G^<_{\mu,\nu}(E)-G^<_{\nu,\mu}(E)]
\label{eq Jmunu}
\end{equation}
can be expressed in terms of the lesser Green's function matrix $G^<$. In noninteracting systems, with the incoming wave sent from one single lead described by self-energy $\Sigma_i$, $G^<$ is given by the kinetic equation $G^<(E) = G^r(E)[\Sigma_i^\dag(E)-\Sigma_i(E)] G^a(E)$, where $G^{r/a}$ is the retarded/advanced Green's function of the scattering region. 

To treat micron-scale graphene samples, we use a scalable tight-binding model \cite{Liu2015}, with a scaling factor $s_f=8$. This scales the lattice spacing to $a \sim 1\unit{nm}$, enabling us to treat (i) the density range of the order of $10^{12}\unit{cm^{-2}}$, typical for experiments using hBN-encapsulated graphene \cite{Dean2010}, and (ii) a sharp \textit{pn} interface of smoothness $\sim 30\unit{nm}\gg a$, a typical thickness of hBN encapsulation layers \cite{Handschin2016,Note2}. Note that the prefactor in Eq.~\eqref{eq Jmunu} containing the Fermi velocity $v_F$ and the unit area $S=3\sqrt{3}a^2/4$ is irrelevant for current density imaging since only dimensionless profiles are shown. In our simulations, the point-like injector diameter will be fixed as $25\unit{nm}$, not too far from the present technical limit \cite{Handschin2015}.

The presented local current density profiles refer to the magnitude $J(x,y)=[J_x^2(x,y)+J_y^2(x,y)]^{1/2}$ of Eq.~\eqref{eq Jvec}, with the Fermi energy set to $E=0$ and the on-site energy profiles obtained from the carrier density profiles described in Ref.\ \onlinecite{Liu2015}. Figure \ref{fig1}(c) highlights the unique characteristic of the generated electron wave pertaining its narrow shape over micron scales. To quantify the high degree of beam collimation, the inset in Fig.~\ref{fig1}(c) shows the angle distribution histogram of the azimuthal angle $\theta = \arg [J_x(x_\mu,y_\mu)+iJ_y(x_\mu,y_\mu)]$ for sites $\mu$ within the white box area (with totally 80800 sites). The angle distribution width is as narrow as $\sim 5^\circ$. Note that the beam generated by a perfect parabolic \textit{pn} junction in the clean limit considered in Fig.~\ref{fig1}(c), as well as the rest of the discussion, is robust against disorder, as long as the mean free path is much longer than the focal length, and practically insensitive to the junction edge roughness, if the latter's length scales are shorter than the Fermi wavelength \footnote{For numerical examples about potential disorder in graphene samples, edge roughness of the parabolic gate, and smoothness of the \textit{pn} junction, and a brief discussion about finite temperature, see \hyperref[sec supp]{Supplemental Material} appended at the end.}. In addition, all calculations consider zero temperature, since the lensing mechanism is not expected to be vulnerable to finite temperatures \cite{Note2}.

The gallery of panels Fig.~\ref{fig1}(d)--\ref{fig1}(h) demonstrates various extraordinary properties of the focused electron beam: In Fig.\ \ref{fig1}(d), an additional barrier (white dashed lines) with density gated to $-n_o$ is considered. The collimated wave tunnels through the barrier almost reflectionlessly, a consequence of Klein tunneling due to normal incidence of the beam. In Fig.~\ref{fig1}(e), an additional potential step (right of the white dashed line) with density gated to $-n_o$ results in a symmetrically and negatively refracted electron beam (injected from a lensing apparatus tilted by $15^\circ$) as clearly visible from the current density; the blue dotted line marks the expected trajectory in the ray optical limit. The collimation persists also in the presence of a weak perpendicular magnetic field, $\mathbf{B} = (0,0,B)$, where ``weak'' means that the resulting cyclotron radius $r_c = \hbar\sqrt{\pi|n_o|}/eB \gg f$. This is clearly seen in Fig.~\ref{fig1}(f), the blue dotted line marking the expected cyclotron trajectory segment. 

\begin{figure}[b]
\includegraphics[width=\columnwidth]{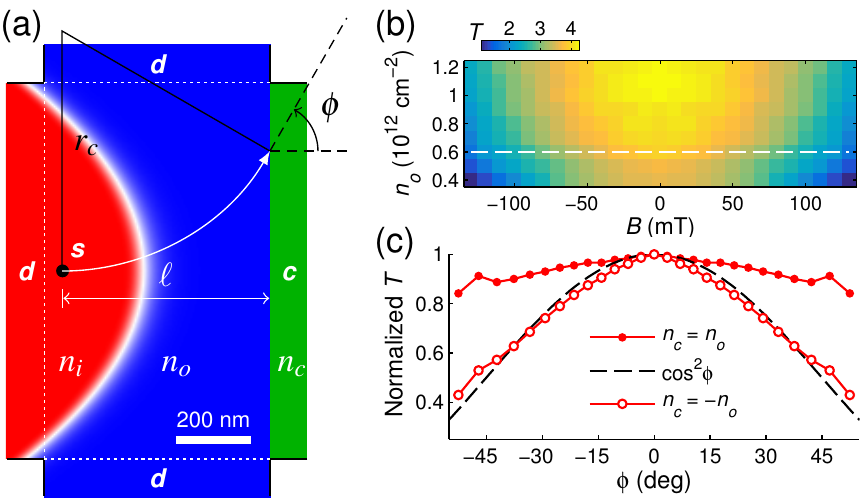}
\caption{(a) Schematic of the lensing apparatus in the presence of a weak magnetic $B$ field. (b) Transmission $T$ for electron flow from the point source ($s$) to the collector ($c$) as a function of field strength and density $n_o$ outside the lens. The density  is set to $n_i=-n_o$ inside the lens, and fixed at $n_c=-6\times 10^{11}\unit{cm^{-2}}$ in $c$. Along the white dashed line, $T(B)$ normalized to its maximum is re-interpreted as $T(\phi)$ in (c), with $\phi(B)$ given by Eq.~\eqref{eq phi}, and compared to $\cos^2\phi$ (black dashed curve). As a reference curve, $T(\phi)$ with $n_c = n_o$ is also shown.\label{fig2}}
\end{figure}

This close correspondence between the quantum mechanical wave propagation and classical cyclotron motion is further illustrated in panels (g) and (h) of Fig.~\ref{fig1}, where blue patches represent transparent semi-infinite leads. The lensing apparatus in the upper left corner of a graphene cavity, shown in panel (g), generates a ``skipping wave'' that tracks a classical skipping orbit, composed of many cyclotron segments along the top, right, and bottom edges that amount to a length of about 10 $\mu m$. Skipping orbits are often considered as the classical analogue of quantum Hall edge channels, albeit in a loose sense. Here, the electron beam represents a particular solution to the Schr\"odinger equation that probably can be regarded as maximally classical, though still subject to interference. Correspondingly, the ring wave mimicking a full cyclotron orbit, depicted in panel (h) where a double-sided parabolic lens is considered, encloses an Aharonov-Bohm flux.

The ability to both generate such narrow beams and to steer their direction through bending in a $B$-field, with high angular resolution and without losing collimation, immediately opens up the possibility to substantially improve two prominent applications in ballistic graphene electronics, to be described in the following.

First, it enables one to accurately measure the angle-resolved transmission of carriers traversing a \textit{pn} junction, {\em i.e.} angle-dependent Klein tunneling, which has remained a long-standing experimental challenge despite some recent efforts \cite{Sutar2012,Rahman2015}. Using the proposed lensing apparatus, the angle of incidence can be continuously varied by tuning the $B$ field, which bends the electron beam. To simulate such an angle-resolved transmission ``experiment'', we perform a transport calculation considering the geometry in Fig.~\ref{fig2}(a). There, the transparent drain leads labeled by $d$ are to suppress boundary effects from the finite-size graphene lattice. After traversing a distance $\ell$ along the parabola axis, a bent trajectory hits the interface under an angle (with respect to its normal)
\begin{equation}
\phi = \arcsin\frac{eB\ell}{\hbar\sqrt{\pi|n_o|}}\ ,
\label{eq phi}
\end{equation}
which can be controlled by the field strength $B$ and density $n_o$.

Figure \ref{fig2}(b) shows the transmission $T$ for charge flow from the source $s$ to the collector $c$ as a function of magnetic field $B$ and density $n_o$ by varying the density inside the lens $n_i = -n_o$ accordingly and fixing $n_c = -6\times 10^{11}\unit{cm^{-2}}$ at the collector. $T(\phi)$ in Fig.~\ref{fig2}(c) is obtained by taking $T(B)$ along the white dashed line cut in Fig.~\ref{fig2}(b) and using $\phi(B)$ given by Eq.~\eqref{eq phi}. Since along this cut the sharp \textit{pn} junction between the scattering region and the collector lead becomes symmetric ($n_o = -n_c$), the transmission function is expected to behave like a cosine squared \cite{Cheianov2006}. As seen in Fig.~\ref{fig2}(c), the normalized $T(\phi)$ indeed agrees well with $\cos^2\phi$. As a reference line, $T(\phi)$ for $n_c=n_o=6\times 10^{11}\unit{cm^{-2}}$ is also shown in Fig.~\ref{fig2}(c), exhibiting a nearly $\phi$-independent form. Both $T(\phi)$ curves with $n_c=-n_o$ and $n_c=n_o$ exhibit a small kink around $\phi\approx \pm 45^\circ$, which is simply a boundary (finite size) effect. By either shortening $\ell$ or increasing the width, it is possible to investigate $T(\phi)$ up to higher angles.

\begin{figure}[b]
\includegraphics[width=\columnwidth]{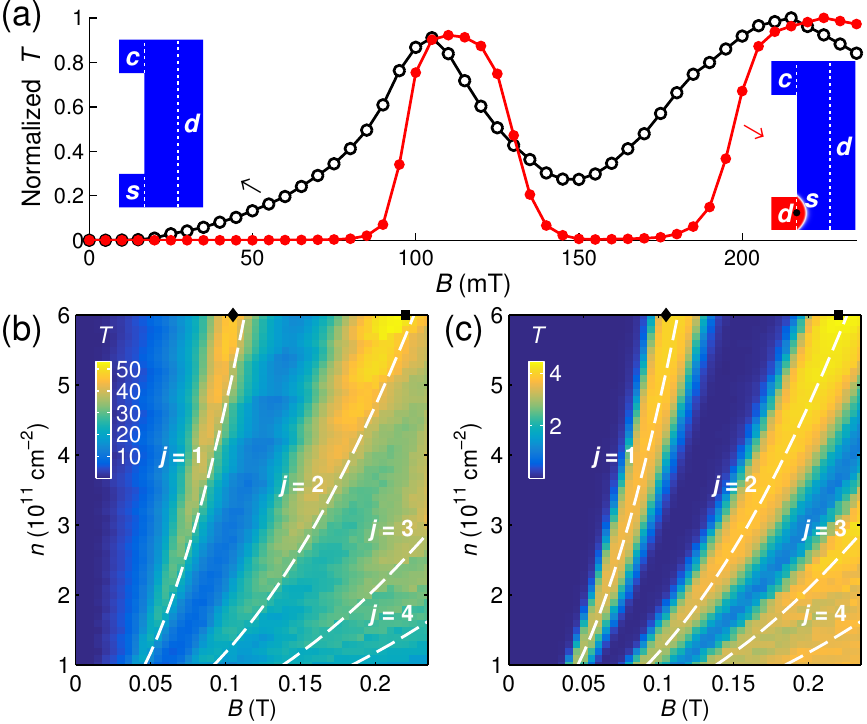}
\caption{(a) Normalized transmission $T$ from source $s$ to collector $c$ as a function of $B$ at density $n=6\times 10^{11}\unit{cm^{-2}}$ in the TMF geometry, without (left inset, black curve) and with (right inset, red curve) the lensing apparatus [similar to Fig.~\ref{fig2}(a)] at lower left terminal. (b)/(c) Color maps of transmission $T(B,n)$ (not normalized) without/with the lensing apparatus. TMF states for $j=1,\ldots ,4$ predicted by Eq.~\eqref{eq TMF} are marked by white dashed lines. Symbols $\blacklozenge$ and $\blacksquare$ in (b) and (c) mark the values of $B$ and $n$ used in Fig.~\ref{fig4}.\label{fig3}}
\end{figure}

Second, controlled bending of the narrow electron beam is also particularly suited to improve transverse magnetic focusing (TMF). Very recently, TMF in high-mobility graphene has gained strong experimental interest \cite{Taychatanapat2013,Calado2014,Morikawa2015,Bhandari2016,Lee2016} as a tool to study and engineer charge carrier flow. TMF requires that the carrier density fulfills
\begin{equation}
n = \frac{1}{\pi}\left(\frac{eB}{h}\frac{D}{j}\right)^2\ ,
\label{eq TMF}
\end{equation}
where $j$ is a positive integer and $D$ is the distance between the midpoints of a source and a collector probe. Here we consider a 2-$\mu$m-wide graphene sample [see left inset in Fig.~\ref{fig3}(a)] with the right side attached to a transparent lead ($d$), such that the sample becomes semi-infinite, and the left side attached to two probes of width $w=0.4\unit{\mu m}$, one source ($s$) and one collector ($c$), separated by $D=1.6\unit{\mu m}$ from each other. We consider only transmission from $s$ to $c$ for a two-point measurement, rather than the six conductance coefficients required for the four-point resistance \cite{Datta1995}. 

For fixed density $n=6\times 10^{11}\unit{cm^{-2}}$, the normalized transmission $T(B)$ is shown by the black curve with open circles in Fig.~\ref{fig3}(a), with two broad peaks corresponding to $j=1$ and $j=2$ in line with Eq.~\eqref{eq TMF}. Replacing the probe $s$ by the lensing apparatus with $f=100\unit{nm}$ [right inset of Fig.~\ref{fig3}(a)], the normalized $T(B)$ is shown by the red curve with solid dots in Fig.~\ref{fig3}(a). The lensing apparatus clearly sharpens the TMF signal by narrowing down the $j=1$ peak width. Most notably, outside the peak, $T(B)$ drops drastically to zero, implying a perfect peak-to-background ratio, as a result of the sharp curved electron beam. In fact, the first TMF peak with lensing occurs roughly between $B=0.1\unit{T}$ and $B=0.13\unit{T}$, corresponding to cyclotron diameters of $2r_c\approx 1.81\unit{\mu m}$ and $2r_c\approx 1.39\unit{\mu m}$, respectively.  The difference $\approx 0.42\unit{\mu m}$ agrees well with the collector probe width of $w=0.4\unit{\mu m}$, again suggesting a highly concentrated electron beam. In Figs.~\ref{fig3}(b)/\ref{fig3}(c), we show $T(B,n)$ color maps without/with the lensing apparatus; the latter clearly exhibits enhanced $j=1,2$ TMF peaks.

\begin{figure}[t]
\includegraphics[width=\columnwidth]{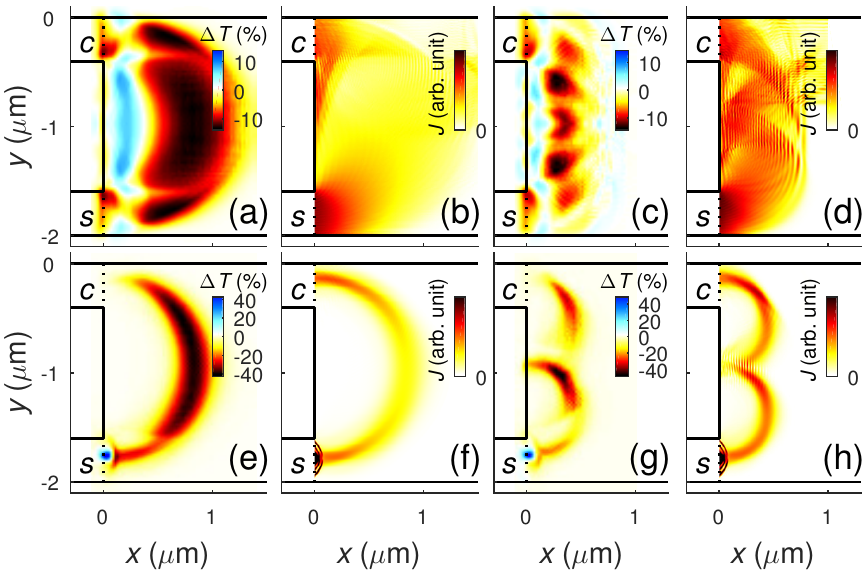}
\caption{Scanning gate images $\Delta T(x,y)$ without/with the lensing apparatus for (a)/(e) $j=1$ and (c)/(g) $j=2$ TMF states, and their correspnding probability current density distribution $J(x,y)$ for (b)/(f) $j=1$ and (d)/(h) $j=2$. Values of magnetic field $B$ and carrier density $n$ used in (a,b,e,f) and (c,d,g,h) correspond to $\blacklozenge$ and $\blacksquare$ marked in Fig.~\ref{fig3}, respectively.\label{fig4}}
\end{figure}

Finally, we consider and simulate scanning gate microscopy (SGM) as a tool to monitor charge carrier flow. In SGM experiments, a capacitively coupled charged tip is scanned over a phase-coherent sample, thus acting as a tunable and movable scatterer, and the sample conductance (or resistance in four-point measurements) is measured as a function of the tip position $\mathbf{r}_\mathrm{tip}$. The difference $\Delta G(\mathbf{r}_\mathrm{tip}) \equiv G(\mathbf{r}_\mathrm{tip})-G_0$ between the sample conductance with ($G$) and without ($G_0$) the tip is plotted as a function of $\mathbf{r}_\mathrm{tip}$. The images thus obtained were originally interpreted as maps of the coherent electron flow through quantum point contacts defined in two-dimensional electron gases (2DEGs) \cite{topinka2001}: Backscattering from the tip in a region where a lot of electrons are passing by will cause a sizable conductance change, the contrary holding true when the tip is positioned away from such ``high flow'' regions. 

Previous theoretical and experimental works considering a variety of phase-coherent systems \cite{Woodside2002,Aidala2007,martins2007,pala2008,jalabert2010,schnez2011,paradiso2012,gorini2013,kozikov2013,Kolasi`nski2014,aoki2014,zhukov2014,kleshchonok2015} showed the versatility of this technique, but also that a general interpretation of an SGM image as a flow map can be problematic \cite{pala2008,jalabert2010,gorini2013,kozikov2013}. In particular, it was shown in Refs.~\onlinecite{jalabert2010,gorini2013} that an explicit connection between local current densities and SGM images requires stringent symmetry conditions. This is consistent with measurements in 2DEGs mesoscopic rings \cite{martins2007,pala2008}, which established a connection between the local density of states and the $\Delta G$ images, as well as with recent theoretical \cite{Kolasi`nski2014} and experimental \cite{kozikov2013} developments.   

In this context, the lensing apparatus is an ideal tool for testing the interpretation of SGM measurements. For the TMF geometry considered in Fig.~\ref{fig3}, we compare in Fig.~\ref{fig4} the calculated SGM images $\Delta T$ and probability current density maps $J(x,y)$, without [Figs.~\ref{fig4}(a)--(d)] and with [Figs.~\ref{fig4}(e)--(h)] the lensing apparatus. Here, $\Delta T(x,y) \equiv [T(x,y)-T_0]/T_0$, where $T_0$ without the perturbing tip has been shown in Figs.~\ref{fig3}(b)--\ref{fig3}(c) and $T(x,y)$ is the transmission function from $s$ to $c$ in the presence of a tip at $\mathbf{r}_\mathrm{tip}=(x,y)$ inducing a local carrier density change modeled by $n_\mathrm{tip}(x,y) = n_\mathrm{tip}^0 h^3(x^2+y^2+h^2)^{-3/2}$ with $n_\mathrm{tip}^0=-5\times 10^{11}\unit{cm}^{-2}$ and $h=50\unit{nm}$ adopted from Ref.~\onlinecite{Bhandari2016}. 

Our three-terminal sample does not meet any particular symmetry requirement, and therefore we do not expect a clear correlation between the local current densities and the SGM maps \cite{gorini2013}. This is confirmed by Figs.~\ref{fig4}(a)--(d): electrons injected into the system generate complex current patterns extending over most of the sample [Figs.~\ref{fig4}(b) and (d)], which are barely reflected by the SGM images [Figs.~\ref{fig4}(a) and (c)] -- note that the latter agree with recent measurements on graphene \cite{Morikawa2015,Bhandari2016}. The lensing apparatus drastically changes the picture. In Figs.~\ref{fig4}(f) and (h) the current densities focus as narrow beams and agree very well with the expected classical trajectories, in sharp contrast to the case without the lensing apparatus. Moreover, the SGM maps in the presence of the lensing apparatus [Figs.~\ref{fig4}(e) and (g)] also show a highly concentrated beam structure that agrees well with the classical trajectories. In other words, the SGM signal and the local current density carry the same information. As a consequence, the system response to the local tip perturbation can be unambiguously interpreted classically in terms of the local current flow. 

In conclusion, we proposed an efficient collimation mechanism to generate narrow, non-dispersive charge carrier beams in graphene, which can be steered by magnetic fields without losing collimation. The lens allows unprecedented control over the electron propagation in ballistic graphene, as demonstrated by the example applications of angle-resolved transmission across a \textit{pn} junction, transverse magnetic focusing, and imaging of the current flow simulating scanning gate microscopy. We expect to excite next-generation graphene electron optics experiments based on the proposed concept for wave collimation. As the underlying mechanism exploits negative refraction and Klein collimation that are unique to pseudo-relativistic Dirac materials, the lensing mechanism may equally apply to surface states of topological insulators.

\begin{acknowledgments}
We thank C.~Handschin, P.~Makk, and P.~Rickhaus for sharing their opinion about experimental feasibility, and R.\ R\"{u}ckner for technical support and system maintenance of Rechenzentrum Universit\"at Regensburg, where most of the presented calculations were performed. Financial support by DFG within SFB 689 is gratefully acknowledged.
\end{acknowledgments}

\bibliographystyle{apsrev4-1}
\bibliography{../../../../mhl2}

\newpage

\section{Supplemental Material}\label{sec supp}

\renewcommand{\thefigure}{S\arabic{figure}}
\renewcommand{\theequation}{S\arabic{equation}}

\appendix

\setcounter{figure}{0}

This supplemental material provides additional numerical examples of the probability current density profiles to 
show the influence of 
\begin{enumerate}[(i)]
\item spatial disorder in graphene samples, 

\item edge roughness of the parabolic gate, and 

\item the smoothness of the \textit{pn} junction 
\end{enumerate}
on the electron beam generated by the lensing apparatus 
described in the main text. 
All numerical examples are meant to be compared with (and are being based on the same parameters as)
the ideal case of Fig.~1(c) in the main text, which considers a focal length $f=200\unit{nm}$, 
densities $n_o=-n_i=6\times 10^{11}\unit{cm^{-2}}$ corresponding to a Fermi wave length 
$\lambda_F\approx 46\unit{nm}$, and a smoothness $\ell_s=30\unit{nm}$ of the \textit{pn} junction profile. 
Moreover, we comment on the expected temperature dependence in the closing.

\begin{figure}[b]
\includegraphics[width=0.9\columnwidth]{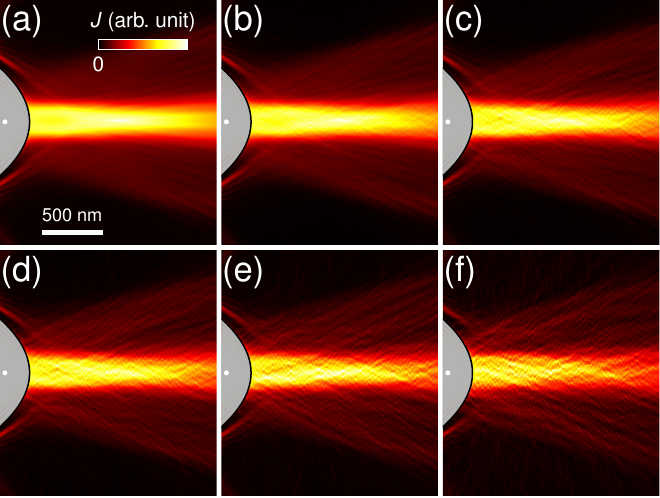}
\caption{Probability current density profiles of the electron beam generated by the lensing apparatus (focal length $f=200\unit{nm}$) in the presence of (a) zero and (b)--(f) finite disorder. The disorder strengths $U_\mathrm{dis}$ in (b), \ldots, (f) are $10,\ldots,50\unit{meV}$. The color in (a) applies to all the $J$ maps in this supplemental material, and so does the scale bar, except in Fig.~\ref{figS2}(b).}
\label{figS1}
\end{figure}

\subsection{Disorder in graphene}

Throughout the main text the graphene lattice is treated as purely ballistic, {\em i.e.} free of disorder, in view of high mobilities for graphene achieved in state-of-the-art experiments with mean free paths $\ell_\mathrm{mfp}$ up to scales beyond $20\unit{\mu m}$ \cite{Banszerus2016}. To test the robustness of the proposed lensing apparatus against disorder, we consider a Hamiltonian with random on-site potential:
\[
H = H_0 + \sum_j U_j c^\dag_j c_j\ .
\]
Here, $H_0$ is the clean part of the scaled tight-binding Hamiltonian \cite{Liu2015} (including the lensing potential), and $U_j\in [-U_\mathrm{dis}/2,U_\mathrm{dis}/2]$ is a random potential at site $j$ fluctuating in a range of $U_\mathrm{dis}$. 

As a reference panel, Fig.~\ref{figS1}(a) shows the beam for the clean system without disorder, repeating Fig.~1(c) of the main text. In panels~\ref{figS1}(b), (c),\ldots, (f) we consider $U_\mathrm{dis}=10\unit{meV}$, $20\unit{meV}$, \ldots, $50\unit{meV}$ which clearly demonstrate the robustness of the beam profile against static potential disorder. Reasonably, the generated electron beam is expected to remain focused in a range shorter than the elastic mean free path $\ell_\textrm{mfp}$. Using Fermi's golden rule, in Fig.~\ref{figS1}(b), (c), (d), (e), and (f) the mean free paths are estimated to be $\ell_\mathrm{mfp}\approx 106.4\unit{\mu m}$, $26.6\unit{\mu m}$, $11.8\unit{\mu m}$, $6.7\unit{\mu m}$, and $4.3\unit{\mu m}$, respectively, covering the current state-of-the-art $\ell_\mathrm{mfp}$ of $28\unit{\mu m}$ \cite{Banszerus2016}. Naturally, the lensing apparatus is expected to fail in strongly disordered graphene with $\ell_\mathrm{mfp}\lesssim f$. Note that the relevant transport mean free path, equal to the elastic mean free path for white-noise type disorder ({\em s}-wave scattering) considered here, will be even longer for long-range potential fluctuations, implying an even more robust beam profile in the latter case.

\subsection{Edge roughness of the parabolic gate}

\begin{figure}[b]
\includegraphics[width=0.9\columnwidth]{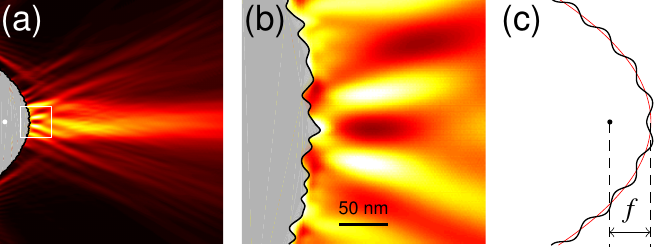}
\caption{(a) Imperfect beam profile considering a strong line edge roughness \cite{Ferry2009,Fang2008,Yazdanpanah2012} (fluctuation range and correlation length $\sim 20\unit{nm}$) modulating the parabolic gate profile. The region marked by a white box is zoomed in and shown in (b). (c) Schematic of a perfect parabola (red thin line) modulated by a cosine function (black thick line) with an amplitude $\delta=f/10$ and period $\lambda=f$.}
\label{figS2}
\end{figure}

The parabolic $pn$ junction in our lensing apparatus can be experimentally realized by a local gate etched in a parabolic shape, which is expected to be imperfect in reality. Depending on the resolution of the $e$-beam lithography and the material properties of the polymer mask used during the fabrication process, the profile of the parabolic gate may exhibit edge roughness. Based on a mathematical model estalblished for studying surface roughness \cite{Ferry2009}, which was later adopted to investigate the effect of line edge roughness (LER) in graphene nanoribbons \cite{Fang2008,Yazdanpanah2012}, an example of the resulting current density $J$ is shown in Fig.~\ref{figS2}(a), considering rather strong LER parameters [see Fig.~\ref{figS2}(b) for the actual edge profile considered in Fig.~\ref{figS2}(a)].

\begin{figure}[b]
\includegraphics[width=\columnwidth]{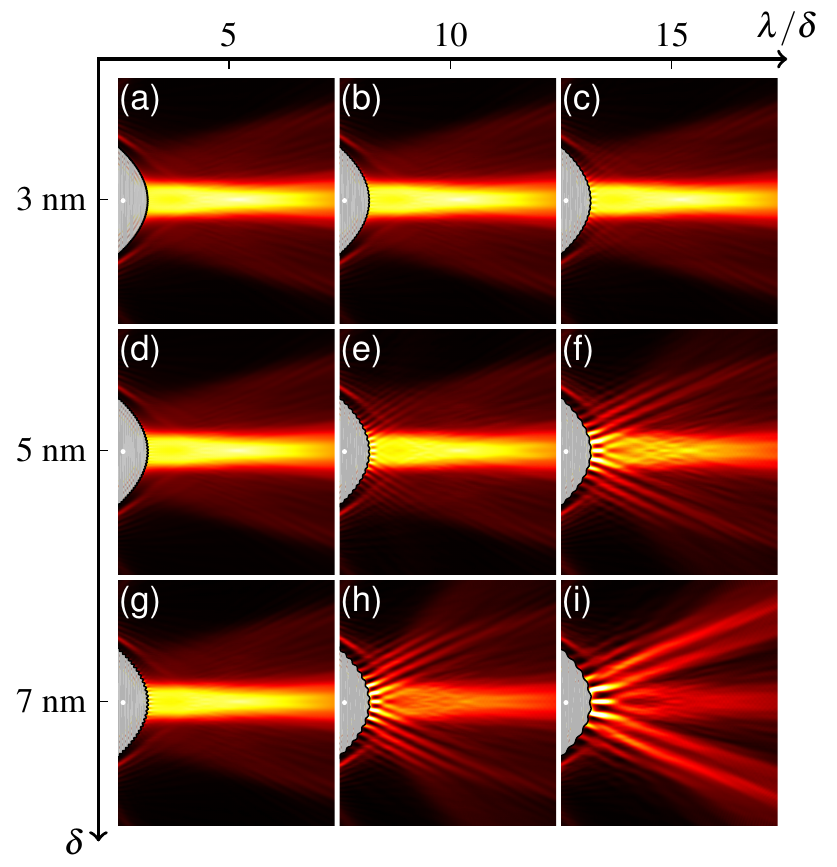}
\caption{Probability current density profiles of the electron beam generated by an imperfect parabolic lens based on the simplified model sketched in Fig.~\ref{figS2}(b) with various roughness amplitudes $\delta$ (left to right panels) and periods $\lambda$ (upper to lower panels).}
\label{figS3}
\end{figure}

To better quantify the effect of the edge roughness, in the following we consider a simplified model with a regular fluctuating profile perpendicular to the parabola described by $\delta\cos(2\pi s/\lambda)$, where $s$ is the arc length of the parabola with respect to its vertex. See an exemplary sketch in Fig.~\ref{figS2}(c). Since the edge roughness may typically fluctuate within a range of $2\delta\sim 10\unit{nm}$ \footnote{C.\ Handschin, P.\ Makk, and P.\ Rickhaus, private communication.}, we consider $\delta=3\unit{nm}$, $5\unit{nm}$, and $7\unit{nm}$ in Fig.~\ref{figS3}, each with various roughness periodicities of $\lambda=5\delta,10\delta,15\delta$ (for $\lambda=\delta$ we do not observe any observable distortion of the beam). Recall from the main text that the Fermi wave length is $\lambda_F\approx 46\unit{nm}$, which is much longer than the considered fluctuation amplitude $\delta$. The parameter $\delta$ is therefore expected to play a minor role. On the other hand, the considered periodicities cover the case $\lambda>\lambda_F$, when the roughness is expected to show pronounced effects.

\autoref{figS3} confirms this intuitive expectation. In panels (a)--(c) with $\delta=3\unit{nm}$ and $\lambda=15,30,45\unit{nm}$, the beam structure remains perfectly focused in all cases, even for $\lambda\approx\lambda_F$ in Fig.~\ref{figS3}(c), where visible leakage fringes in the vicinity of the modulated parabolic junction in addition to the collimated electron beam can be observed. In Figs.~\ref{figS3}(d)--(f) with $\delta=5\unit{nm}$, similar patterns with nearly perfect beam structures for $\lambda=25\unit{nm}$ (d) and $\lambda=50\unit{nm}$ (e) are observed, while for $\lambda=75\unit{nm}>\lambda_F$ (f) significant leakage of secondary beams can be seen. Such secondary beams follow the wavy structure of the modulated parabolic $pn$ junction, and can be observed in Figs.~\ref{figS3}(h) and (i), both with $\lambda > \lambda_F$. Despite the large fluctuation amplitude of $\delta=7\unit{nm}$ that already exceeds the typical roughness of $2\delta\sim 10\unit{nm}$ \cite{Note1}, the electron beam shown in Fig.~\ref{figS3}(g) with $\lambda=35\unit{nm}<\lambda_F$ remains nearly perfect.

We conclude that the electron beam generated by the proposed lensing apparatus is practically insensitive to the possible edge roughness of the parabolic gate, as long as the fluctuation amplitude and correlation length of the roughness are both smaller than the Fermi wave length.

\subsection{Smoothness of the \textit{pn} junction}

In the main text, as well as in Figs.~\ref{figS1}--\ref{figS3}, the carrier density function modeling the parabolic junction has been described by a smoothness of $\ell_s=30\unit{nm}$ bridging the inner and outer densities. In experiments, the length scale of this smoothness is determined by the distance from the graphene sample to the parabolic gate. The idea of the lensing apparatus was motivated by the experimental work of \cite{Handschin2015} for point contacts in hBN-encapsulated graphene (hBN stands for hexagonal boron nitride), implying $\ell_s\sim t_\mathrm{hBN}$. Here $t_\mathrm{hBN}$ is the thickness of the hBN encapsulation layer separating the graphene sample and the gate electrode, and ranges typically from several to a few tens of nm. As explained in a technical remark in the main text, the chosen $\ell_s=30\unit{nm}$ is not only a typical hBN layer thickness but also well satisfies $a\ll\ell_s$, where $a$ is the scaled lattice spacing, required for the scalable tight-binding model (TBM) \cite{Liu2015} to be precise enough.

\begin{figure}[b]
\includegraphics[width=\columnwidth]{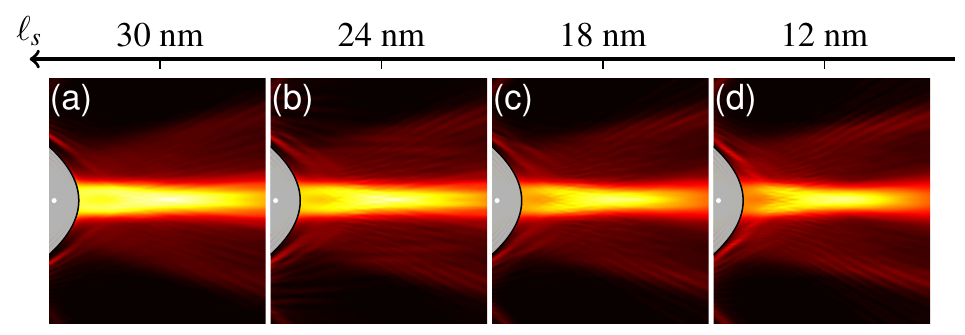}
\caption{Probability current density profiles of the electron beam generated by a parabolic $pn$ junction with varying smoothness. Panel (a) with smoothness $\ell_s=30\unit{nm}$, identical to Fig.~\ref{figS1}(a), is shown for reference; panels (b)--(d) display results for  sharper junctions with smaller $\ell_s =24,18,12\unit{nm}$.}
\label{figS4}
\end{figure}

Specifically, our lattice spacing scaled by a factor $s_f=8$ is $a\approx 1\unit{nm}$, corresponding to $1/30$ of the smoothness for $\ell_s=30\unit{nm}$ and thereby ensuring a high precision of the scalable TBM. In Fig.~\ref{figS4}, the $J$ profiles of the generated beam considering smoothness of $\ell_s=24,18,12\unit{nm}$, in addition to the reference case with $\ell=30\unit{nm}$, are shown. Clearly, the beam structure exhibits only minimal changes for reduced smoothness. The sharpest junction considered in Fig.~\ref{figS4}(d) has $a\approx \ell_s/12 \ll \ell_s$ so that the scalable TBM should work well in all panels of Fig.~\ref{figS4}, however we note that its slight decrease in precision from (a) to (d) may be the cause of the fine structure outside the beam vaguely visible, in particular in panel (d). 

Overall, the beam structure is shown to be rather insensitive to the smoothness of the parabolic \textit{pn} junction for typical hBN layer thicknesses sandwiching the graphene sample.

\subsection{Temperature dependence}

All calculations here and in the main text were done at the Fermi level and assuming zero temperature. 
This is a very good approximation, since most ballistic graphene transport experiments are typically performed
at temperatures around $1\unit{K}$. Still, we comment on possible effects on our lensing apparatus due to finite 
temperature.

\paragraph{Thermal broadening.} 

As the energy scale of thermal broadening is around $\Delta E_\mathrm{th}\sim 10^{-4}\unit{eV}$ at a temperature $T=1\unit{K}$, negligible compared to typical transport energy scales in graphene \textit{pn} junctions, we expect the lensing apparatus to work up to a few tens of Kelvin. Specifically, the densities $n_o=-n_i=6\times 10^{11}\unit{cm^{-2}}$ considered in our numerical examples correspond to a Fermi energy $E_F\approx 90\unit{meV}$. Thus for the thermal broadening to be appreciable, $\Delta E_\mathrm{th}\sim E_F/10$, which corresponds to a temperature around $100\unit{K}$.

\paragraph{Increased scattering rate.} 

With increasing temperature the inelastic scattering rate due to electron-electron and electron-phonon interactions increases, and hence, vice versa, the effecive electron mean free path $\ell_\mathrm{mfp}$ decreases once $T$-dependent inelastic scattering exceeds elastic potential scattering. This is the basic mechanism that e.g.\ suppresses the Fabry-P\'erot-type interference observed in graphene \textit{pnp} cavities of length $\ell_c$ when increasing the temperature from a few $\unit{K}$ (at which $\ell_\mathrm{mfp}>\ell_c$) to a few tens of $\unit{K}$ (at which $\ell_\mathrm{mfp}<\ell_c$); see, for example, \cite{Young2009,Grushina2013}. As shown in one of the state-of-the-art works on hBN-encapsulated graphene \cite{Wang2013}, $\ell_\mathrm{mfp}$ indeed decreases with $T$ exponentially but still remains several microns at temperatures as high as $T=100\unit{K}$. Thus the proposed lensing apparatus, if carried out in such a high-quality sample, would, in principle, persist up to such temperatures. The focal length $f$ sets the mimimum scale for $\ell_\mathrm{mfp}$ required such that a collimated beam can be formed by the lensing apparatus.
 
To conclude the above discussion, the lensing mechanism proposed in the main text is not expected to be vulnerable to finite temperature. Indeed, in high-quality samples as \cite{Wang2013} at energy scales around $0.1\unit{eV}$ or above, it is expected to work at temperature $T\lesssim 100\unit{K}$.

\end{document}